\documentclass[aps,prd,twocolumn,nofootinbib]{revtex4-1}
\usepackage{epsfig}
\usepackage[colorlinks,linkcolor=blue,anchorcolor=blue,citecolor=blue,urlcolor=blue,breaklinks=true]{hyperref}
\usepackage{graphics}
\usepackage{color}

\begin{document}
\author{Shu-Sheng Xu$^{1,7}$, Zhu-Fang Cui$^{2,7}$, Bin Wang$^{3}$, Yuan-Mei Shi$^{4}$, You-Chang Yang$^{2,5}$, and Hong-Shi Zong$^{2,6,7,}$}\email[]{Email: zonghs@nju.edu.cn}
\address{$^{1}$ Key Laboratory of Modern Acoustics, MOE, Institute of Acoustics, and Department of Physics, Nanjing University, Nanjing 210093, China}
\address{$^{2}$ Department of Physics, Nanjing University, Nanjing 210093, China}
\address{$^{3}$ Department of Physics, Huazhong University of Science and Technology, Wuhan 430074, China}
\address{$^{4}$ Department of Physics and Electron ic Engineering, Nanjing Xiaozhuang University, Nanjing 211171, China}
\address{$^{5}$ School of Physics and Mechanical-Electrical Engineering, Zunyi Normal College, Zunyi 563002, China}
\address{$^{6}$ Joint Center for Particle, Nuclear Physics and Cosmology, Nanjing 210093, China}
\address{$^{7}$ State Key Laboratory of Theoretical Physics, Institute of Theoretical Physics, CAS, Beijing, 100190, China}

\title{The chiral phase transition with a chiral chemical potential in the framework of Dyson-Schwinger equations}
\begin{abstract}
Within the framework of Dyson-Schwinger equations (DSEs), we discuss the chiral phase transition of QCD with a chiral chemical potential $\mu_5$ as an additional scale. We focus especially on the issues related to the widely accepted as well as interested critical end point (CEP). With the help of a scalar susceptibility, we find that there might be no CEP$_5$ in the $T-\mu_5$ plane, and the phase transition in the $T-\mu_5$ plane might be totally crossover when $\mu<50$ MeV, which has apparent consistency with the Lattice QCD calculation. Our study may also provide some useful hints to some other studies related to $\mu_5$.

\bigskip


\bigskip

\noindent PACS Number(s): 11.30.Rd, 25.75.Nq, 12.38.Mh, 12.39.-x

\end{abstract}
\maketitle

\section{Introduction}\label{intro}
Quantum Chromodynamics (QCD), which describes the interactions between quarks and gluons, is already commonly accepted as an essential part of the Standard Model of particle physics. Dynamical chiral symmetry breaking (DCSB) and quark color confinement are two fundamental features of QCD, and there are also many laboratories and experiments on this field, such as the famous Relativistic Heavy Ion Collider (RHIC) and Large Hadron Collider (LHC). However, thanks to the complicated non-Abelian feature of QCD itself, it is so difficult to have a thorough understanding of the mechanisms of DCSB and confinement, especially in the interesting non-perturbative region, which means quarks and gluons are strongly coupled to each other and then the related processes have small momentum transfer (or equivalently, the coupling constant becomes large and running). In this case, nowadays people often and in some sense have to resort to various effective models to study them phenomenologically, such as the chiral perturbation theory~\cite{AoP.158.142--210,NPB.250.465--516,RPP.58.563,PPNP.35.1--80,PhysRevC.80.034909}, the global color symmetry model (GCM)~\cite{AoP.188.20--60,PPNP.39.117--199,PRC.58.1195,PRC.60.055208,PRD.67.074004}, the quasiparticle model~\cite{ZPC.57.671--675,PLB.337.235--239,PRD.52.5206,PRC.61.045203,PLB.711.65--70,PhysRevD.85.045009,PhysRevD.86.114028,EPJC.73.2626}, the QCD sum rules~\cite{PR.127.1--97,NPB.279.785--803,PRC.46.R34,NPA.624.527--563}, the Nambu--Jona-Lasinio (NJL) model and the related Polyakov-loop-extended Nambu--Jona-Lasinio (PNJL) model~\cite{PPNP.27.195,RMP.64.649,PR.247.221,PLB.591.277--284,PR.407.205,PRD.73.014019,TEPJC.49.213--217,PRD.77.114028,PLB.662.26,EPJC.73.2612,CUIJMP,EPJC.74.2782}, Lattice QCD~\cite{JHEP.04.050,JHEP.09.073,PhysRevLett.110.172001}, and the Dyson-Schwinger equations (DSEs)~\cite{PPNP.33.477,PPNP.45.S1,IJMPE.12.297,PPNP.61.50,JHEP.04.14,PPNP.77.1-69,JHEP.07.014,PhysRevD.90.114031}. Through these studies, people hope to get profound insight of our nature as well as the early Universe.

Generally speaking, chiral symmetry is an exact global symmetry only when the current quark mass $m$ is zero (the chiral limit case). In the low temperature ($T$) and low chemical potential ($\mu$) phase (the hadronic phase, often referred to as Nambu-Goldstone phase or Nambu phase), this symmetry is spontaneously broken, and as a consequence there exist $N^2_f-1$ ($N_f$ is the number of flavour) pseudoscalar Nambu-Goldstone bosons, meanwhile the QCD vacuum hosts a chiral condensate (two quark condensate) $\langle{\bar q}q \rangle$ (which can actually act as an order parameter for chiral phase transition). At present, it is commonly accepted that when temperature and/or quark chemical potential are high enough, the strongly interacting hadronic matter will undergo a phase transition to some new phase, where the chiral symmetry is restored for the chiral limit case or partially restored for the $m\neq0$ case. This new phase is usually called Wigner phase, and in some sense is related to the famous quark gluon plasma (QGP), which is expected to appear in the ultra-relativistic heavy ion collisions or the inner core of compact stars. As for the nature of the chiral phase transition when $m\neq0$, a popular scenario favors a crossover at small chemical potential, and then turning into a first order chiral transition for larger chemical potential at a critical end point (CEP)~\cite{Pos2006.024}. Such a picture is consistent with most Lattice QCD simulations and various QCD-inspired models, as listed in the last paragraph, however, it is not yet clarified directly from the first principles of QCD. The search for such a CEP is also one of the main goals in the high energy physics experiments, such as the beam energy scan (BES) program at RHIC~\cite{JPG.38.124023,NPA.862.125--131,NPA.904.256c--263c,NPA.904.903c--906c}. Unfortunately, Lattice Monte Carlo simulations cannot be used to resolve this issue directly due to the ``sign problem''~\cite{PhysRevD.75.116003,PLB.682.240--245,PhysRevLett.107.031601,PhysRevD.86.074506,PhysRevD.86.074510,PhysRevD.88.031502}, and until now there is still no firm theoretical evidence for the existence of such a CEP, so the calculations based on some effective QCD models are also irreplaceable nowadays.

In Ref.~\cite{PhysRevD.78.074033}, K. Fukushima $et~al$ firstly introduce the chiral chemical potential (also called axial chemical potential in some other literatures, such as Ref.~\cite{PhysRevD.85.054013}), $\mu_5$, which is conjugated to chiral charge density; and in Ref.~\cite{PhysRevD.84.014011}, M. Ruggieri suggests that the CEP of the chiral phase diagram can be detected by means of Lattice QCD simulations of grand-canonical ensembles with this chiral chemical potential. By concrete calculations within some chiral models, the author shows that a continuation of the CEP at finite temperature and finite chemical potential, to a possible ``CEP$_5$" in the $T-\mu_5$ plane is reachable, which is then helpful in the determining of the CEP in the $T-\mu$ plane from Lattice QCD. The existence of such a possible CEP$_5$ is also confirmed in some other chiral model studies, for example, Refs.~\cite{PhysRevD.81.114031,PhysRevD.83.105008}. In Ref.~\cite{PhysRevD.85.054013}, the authors investigated the effect of the vector interaction as well as the finite current quark mass on the location of the CEP. In this paper, we will discuss the related topics within the framework of Dyson-Schwinger equations, which is widely used as well as has been proved to be successful in hadron physics and phase transitions of strongly interacting matters. The following of this paper is organized in such a way: in Sect.~\ref{dsesandgp} we give a basic introduction to the DSEs at finite temperature and nonzero chemical potential as well as an effective model gluon propagator, and with the help of a scalar susceptibility we also discuss the nature of the chiral phase transition within this framework; then in Sect.~\ref{chiralmu}, we discuss the influences of the chiral chemical potential on the chiral phase transition of QCD in detail, and mostly focus on the behaviours of the CEP, not only the algebra but also the numerical results; at last, a brief summary is given in Sect.~\ref{sum}.

\section{Dyson Schwinger equations and an effective gluon propagator}\label{dsesandgp}
In this section, we will briefly review the formula of Dyson Schwinger equations, which is not only widely used in the non-perturbative region of QCD, but also in some other fields like the Quantum Electrodynamics in (2+1) dimensions (QED$_3$)~\cite{PhysRevD.29.2423,PPNP.33.477,PhysRevD.90.036007,PhysRevD.90.065005,PhysRevD.90.073013}, etc. At zero temperature and zero chemical potential, the DSE for the quark propagator reads~\cite{PPNP.33.477} (we will always work in Euclidean space and take the number of flavors $N_f=2$ while the number of colors $N_c=3$ throughout this paper. Moreover, as we employ a ultra-violet finite model, renormalization is actually unnecessary)
\begin{equation}
S(p)^{-1}=S_0(p)^{-1}+\frac{4}{3} \int\frac{\mathrm{d}^4q}{(2\pi)^4}g^2D_{\mu\nu}(p-q)\gamma_\mu S(q)\Gamma_\nu,\label{dse1}
\end{equation}
where $S(p)$ is the dressed quark propagator,
\begin{equation}
S_0(p)^{-1}=i\gamma\cdot p+m,
\end{equation}
is the inverse of the free quark propagator, $g$ is the strong coupling constant, $D_{\mu\nu}(p-q)$ is the dressed gluon propagator, and $\Gamma_\nu=\Gamma_\nu(p,q)$ is the dressed quark-gluon vertex. According to the Lorentz structure analysis, we have
\begin{equation}
S(p)^{-1}=i{\not\!p}A(p^2)+B(p^2),\label{ppg1}
\end{equation}
where $A(p^2)$ and $B(p^2)$ are scalar functions of $p^2$. After the gluon propagator together with the quark-gluon vertex are specified, people can then solve this equation numerically.

The extension of the above quark DSE to its nonzero temperature and nonzero quark chemical potential version is systematically accomplished by transcription of the quark four-momentum via $p\rightarrow p_k=(\vec{p}, \tilde\omega_k)$, where $\tilde\omega_k=\omega_k+i\mu$ with $\omega_k=(2k+1)\pi T$, $k\in Z\!\!\!\!Z$ the fermion Matsubara frequencies, and no new parameters are introduced~\cite{PPNP.45.S1}
\begin{eqnarray}\label{dse2}
S(p_k)^{-1}=S_0(p_k)^{-1}+\frac{4}{3}T \int\!\!\!\!\!\!\!\!\sum g^2D_{\mu\nu}(p_k-q_n)\gamma_{\mu}S(q_n)\Gamma_{\nu}.\nonumber\\
\end{eqnarray}
where
\begin{equation}\label{free}
S_0(p_k)^{-1}=i\gamma \cdot p_k+m,
\end{equation}
and $\int\!\!\!\!\!\!\!\sum$ denotes $\sum_{l=-\infty}^{+\infty}\int\frac{\mathrm{d}^3\vec{q}}{(2\pi)^3}$. Nevertheless, its solution now should have four independent amplitudes due to the breaking of $O(4)$ symmetry down to $O(3)$ symmetry~\cite{PPNP.45.S1}
\begin{eqnarray}
S(p_k)^{-1}=&&i\not\!\vec{p}\,A(\vec{p}\,^2,\tilde\omega_k^2) + \mathbf{1}B(\vec{p}\,^2,\tilde\omega_k^2)\nonumber\\
 &&+i\gamma_4\,\tilde\omega_kC(\vec{p}\,^2,\tilde\omega_k^2)+\not\!\vec{p}\,\gamma_4\,\tilde\omega_kD(\vec{p}\,^2,\tilde\omega_k^2),
\end{eqnarray}
where $\not\!\!\vec{p}=\vec{\gamma}\cdot\vec{p}$, $\vec{\gamma}=(\gamma_1, \gamma_2, \gamma_3)$, and the four scalar functions ${\cal F}= A$, $B$, $C$, $D$ are complex and satisfy
\begin{equation}
{\cal F}(\vec{p}\,^2,\tilde\omega_k^2)^\ast ={\cal F}(\vec{p}\,^2,\tilde\omega_{-k-1}^2)\,.\label{abc}
\end{equation}
But as discussed in Ref.~\cite{PPNP.45.S1}, the dressing function $D$ is power-law suppressed in the ultra-violate region, so that actually does not contribute in all cases investigated in our work. At zero temperature but nonzero chemical potential case, $D$ vanishes exactly since the corresponding tensor structure has the wrong transformation properties under time reversal~\cite{ZPA.352.345--350}. For these reasons, in most cases we can just neglect $D$, and get the commonly used general structure of the inverse of quark propagator as
\begin{equation}
S(p_k)^{-1}=i\not\!\vec{p}\,A(\vec{p}\,^2,\tilde\omega_k^2) + \mathbf{1}B(\vec{p}\,^2,\tilde\omega_k^2)+i\gamma_4\,\tilde\omega_kC(\vec{p}\,^2,\tilde\omega_k^2).\label{ppg3}
\end{equation}

For the dressed-gluon propagator, the general form is like this,
\begin{equation}\label{mtgluon}
g^2D_{\mu\nu}(k_{nl})=P_{\mu\nu}^TD_T(\vec{k}^2,\omega_{nl}^2)+P_{\mu\nu}^LD_L(\vec{k}^2,\omega_{nl}^2),
\end{equation}
where $k_{nl}=(\vec{k},\omega_{nl})=(\vec{p}-\vec{q},\omega_n-\omega_l)$, $P_{\mu\nu}^{T,L}$ are transverse and longitudinal projection operators, respectively. And for the domain $T<0.2$ GeV, of which we are concerned in this work, the authors of Ref.~\cite{PhysRevD.75.076003} have proved that $D_T=D_L$ is a good approximation. For the in-vacuum interaction, in this work we will adopt the following form of $Ansatz$ as in Ref.~\cite{PRL.106.172301},
\begin{equation}\label{simMT}
D_T=D_L=D_0\frac{4\pi^2}{\sigma^6}k_{nl}^2e^{-k_{nl}^2/\sigma^2},
\end{equation}
which is a simplified version of the famous as well as widely used one in Refs. \cite{PhysRevC.56.3369,PhysRevC.60.055214}.
\textcolor{blue}{It can be proved that this dressed gluon propagator at $T=0$ violates the axiom of reflection positivity \cite{glimm1987quantum}, and is therefore not observable; i.e., the excitation it
describes is confined. The same is true of the dressed quark propagator which is also not positive definite and hence is confined (Actually, we can take the gluon propagator as input, and the quark
propagator can then be solved numerically. The results show that there is no singularity on the real, positive, i.e. timelike, $p^2$ axis, which implies that quarks are confined).}

As concerning the quark-gluon vertex, in this work we will take the rainbow truncation, which means a simple but symmetry-preserving bare vertex is adopted,
\begin{equation}\label{rainbow}
\Gamma_\nu(p_n,q_l)=\gamma_\nu.
\end{equation}
The status of propagator and vertex studies can be tracked from Ref.~\cite{PPNP.61.50--65}.

\begin{figure}
\includegraphics[width=0.45\textwidth]{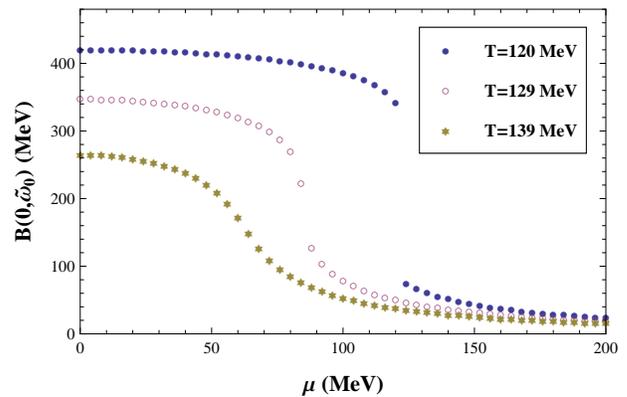}
\caption{$B(0,\tilde\omega_0^2)$ as a function of $\mu$ for three different $T$.}\label{bmu}
\end{figure}

\begin{figure}
\includegraphics[width=0.45\textwidth]{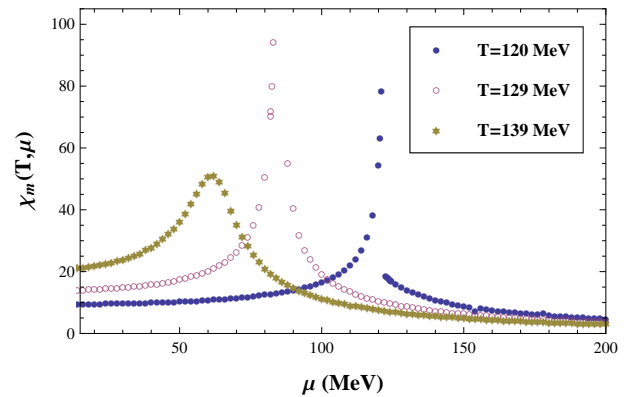}
\caption{$\chi_m(T,\mu)$ as a function of $\mu$ for three different $T$.}\label{sus}
\end{figure}

Now let us fix the related parameters and then show some of the numerical results. $D_0$ and $\sigma$ are usually fixed by fitting the observables, such as the two-quark condensate, the pion decay constant ($f_\pi=131$ MeV) and the pion mass ($m_\pi=138$ MeV). In this work we adopt the one from Ref.~\cite{PhysRevC.60.055214}, that $D_0=9.3\times10^5$ MeV$^2$ and $\sigma=400$ MeV. For the current quark mass we will use $m$=5 MeV. Then substituting Eqs.~(\ref{free}), (\ref{ppg3}), (\ref{simMT}), and (\ref{rainbow}) into Eq.~(\ref{dse2}), we can solve the quark DSE for each value of temperature and chemical potential by means of numerical iteration. As an example, we show $B(0,\tilde\omega_0^2)$ as a function of $\mu$ for different $T$ in Fig.~\ref{bmu}, and the corresponding chiral susceptibility with respect to $m$, which is defined as
\begin{equation}\label{chim}
\chi_m(T,\mu)=\frac{\partial B(0,\tilde\omega_0^2)}{\partial m},
\end{equation}
in Fig.~\ref{sus}.

In general, we can see from Fig.~\ref{bmu} that the scalar function $B(0,\tilde\omega_0^2)$ will decrease when the chemical potential $\mu$ increase, this phenomenon holds to be true for the temperature $T$ and momentum $\vec{p}^2$ too. It is known that the scalar part $B(\vec{p}\,^2,\tilde\omega_k^2)$ of the quark propagator Eq.~(\ref{dse2}) in some sense reflects the dressing effect of the quark, so the results show that the dressing effect becomes weaker and weaker for higher $T, \mu$ and $\vec{p}^2$. We can also see from Fig.~\ref{bmu} that for different values of $T$, $B(0,\tilde\omega_0^2)$ may behave different: for $T$ larger than a critical $T_c=129$ MeV, $B(0,\tilde\omega_0^2)$ change gradually but continuously from the Nambu solution to the Wigner solution; while for $T$ smaller than $T_c$, there will appear a sudden discontinuity at some critical $\mu$.

To study the nature of the chiral phase transition, especially to determine the critical value of $\mu_c$ at $T_c$, people often employ various susceptibilities of QCD~{\cite{PRC.72.035202,PRC.73.035206,PLB.639.248--257,PLB.669.327--330,PRC.79.035209}. We can see from Fig.~\ref{sus} that, for $T\geq T_c$, the susceptibility $\chi_m$ indicate a crossover from the Nambu phase to the Wigner phase, and the peak grow higher and higher when $T$ approaches $T_c$. At $T_c$, $\chi_m$ shows a sharp and narrow divergent peak, and the value of this peak turns to be $\infty$, which demonstrate that here is a second-order phase transition, and corresponding to the CEP. And for $T\leq T_c$, an obvious first order phase transition will occur. According to these results, we can move on to study the chiral phase transition, especially the behavior of CEP. In the following Sec.~\ref{chiralmu}, we will focus on the variance of CEP when the chiral chemical potential is considered as an additional scale.

\section{Influences of the chiral chemical potential on the chiral phase transition}\label{chiralmu}
The concept of chiral chemical potential was first proposed by K. Fukushima $et~al.$ in a study related to the external magnetic field~\cite{PhysRevD.78.074033}. Since topological charge changing transitions can induce an asymmetry between the number of right- and left-handed quarks due to the axial anomaly, they introduce the chiral chemical potential $\mu_5$, which couples to the difference between the number of right- and left-handed fermions. The chirality is also expected to be produced in the high temperature phase of QCD~\cite{PhysRevD.85.054013}. Many researchers argue that although $\mu_5$ is a mere mathematical artifice instead of a true chemical potential\footnote{The reason for this is easy to understand, since the difference in densities of the right- and left-handed quarks, $n_5=n_R-n_L$, does not conserve.}, it has the advantage that can be simulated on the Lattice QCD with $N_c=3$, hence is likely to provide some useful information for the studies of the CEP, and even for inhomogeneous phases or the inner structure of compact stellar objects. One of the most interesting features of the introduction of $\mu_5$ is that, it makes the continuation of the CEP to a possible ``CEP$_5$'' in the $T-\mu_5$ plane possible, which is of course helpful in the determining of the CEP from Lattice QCD. Some other researchers also confirm such a possible CEP$_5$ in related chiral model studies~\cite{PhysRevD.81.114031,PhysRevD.83.105008}.
In this part, we will discuss the related topics within the framework of Dyson-Schwinger equations.

To be specific, in order to study the effects of $\mu_5$, people should add the following term to the Lagrangian density,
\begin{equation}\label{lagrmu5}
\mu_5\bar{\psi}\gamma_4\gamma_5\psi.
\end{equation}
And in our work, the quark propagator and its inverse now can include at most eight components according to the Lorentz structure analysis, namely,
\begin{equation}\label{lorentz}
1, \not\!\vec{p}, \gamma_4, \not\!\vec{p}\gamma_4, \gamma_5, \not\!\vec{p}\gamma_5, \gamma_4\gamma_5, \not\!\vec{p}\gamma_4\gamma_5.
\end{equation}
So now the general inverse form of the dressed quark propagator is
\begin{eqnarray}\label{S-1}
&&S(p_n,\mu_5)^{-1}=i\not\!\vec{p}A+B+i\gamma_4\tilde\omega_nC+\not\!\vec{p}\gamma_4\tilde\omega_nD\nonumber\\
&&+(i\not\!\vec{p}A_5+B_5+i\gamma_4\tilde\omega_nC_5+\not\!\vec{p}\gamma_4\tilde\omega_nD_5)\gamma_5,
\end{eqnarray}
The eight scalar functions ${\cal F}= A$, $B$, $C$, $D$, $A_5$, $B_5$, $C_5$, $D_5$ denotes ${\cal F}={\cal F}(\vec{p}^2,\tilde\omega_n^2,\mu_5)$, which are all complex and satisfy the following equation,
\begin{equation}
{\cal F}(\vec{p}\,^2,\tilde\omega_k^2,\mu_5)^\ast ={\cal F}(\vec{p}\,^2,\tilde\omega_{-k-1}^2,\mu_5)\,.\label{FABC}
\end{equation}

Now the quark DSE at nonzero temperature and nonzero chemical potential is then,
\begin{eqnarray}\label{dse}
&&S(p_n,\mu_5)^{-1}=S_0(p_n,\mu_5)^{-1}\nonumber\\
&&+\frac{4}{3}T\int\!\!\!\!\!\!\!\!\sum g^2D_{\mu\nu}(p_n-q_l)\gamma_\mu S(q_l,\mu_5)\Gamma_\nu(p_n,q_l),
\end{eqnarray}
where
\begin{equation}
S_0(p_n,\mu_5)^{-1}=i\not\!\vec{p}+m+i\gamma_4\tilde\omega_n-\mu_5\gamma_4\gamma_5.
\end{equation}
For the details of $S(q_l,\mu_5)$, please see the Appendix part.

Substituting Eqs.~(\ref{S-1}), (\ref{mtgluon}), and (\ref{rainbow}) into Eq.~(\ref{dse}), we found that the solution is the following coupled integral equations (for the sake of concise, here all the notion ${\cal F}(p)$ means ${\cal F}(\vec{p}\,^2,\tilde\omega_n^2,\mu_5)$)
\begin{eqnarray}\label{Aetal}
A(p)&=&1+c(T)\int\!\!\!\!\!\!\!\!\sum k_{nl}^2e^{-k_{nl}^2/\sigma^2}\times \mathcal{K}_A,\nonumber\\
B(p)&=&m+c(T)\int\!\!\!\!\!\!\!\!\sum k_{nl}^2e^{-k_{nl}^2/\sigma^2}\times \mathcal{K}_B,\nonumber\\
C(p)&=&1+c(T)\int\!\!\!\!\!\!\!\!\sum k_{nl}^2e^{-k_{nl}^2/\sigma^2}\times \mathcal{K}_C,\nonumber\\
D(p)&=&c(T)\int\!\!\!\!\!\!\!\!\sum k_{nl}^2e^{-k_{nl}^2/\sigma^2}\times \mathcal{K}_D,\nonumber\\
A_5(p)&=&c(T)\int\!\!\!\!\!\!\!\!\sum k_{nl}^2e^{-k_{nl}^2/\sigma^2}\times \mathcal{K}_{A_5},\nonumber\\
B_5(p)&=&c(T)\int\!\!\!\!\!\!\!\!\sum k_{nl}^2e^{-k_{nl}^2/\sigma^2}\times \mathcal{K}_{B_5},\nonumber\\
C_5(p)&=&i\mu_5/\omega_nc(T)\int\!\!\!\!\!\!\!\!\sum k_{nl}^2e^{-k_{nl}^2/\sigma^2}\times \mathcal{K}_{C_5},\nonumber\\
D_5(p)&=&c(T)\int\!\!\!\!\!\!\!\!\sum k_{nl}^2e^{-k_{nl}^2/\sigma^2}\times \mathcal{K}_{D_5},
\end{eqnarray}
in which
\begin{eqnarray}\label{aaetal}
c(T)&=&\frac{16\pi^2T}{3\sigma^6},\nonumber\\
\mathcal{K}_A&=&-[(\vec{p}\cdot \vec{q}\, k_{nl}^2+2\vec{k}\cdot \vec{p}\, \vec{k}\cdot\vec{q})\sigma_A+2\vec{k}\cdot\vec{p}\,\omega_{nl}\omega_l\sigma_C]/(\vec{p}^2 k_{nl}^2),\nonumber\\
\mathcal{K}_B&=&3\sigma_B,\nonumber\\
\mathcal{K}_C&=&-\omega_l\sigma_C/\omega_n-2\omega_{nl}(\vec{k}\cdot\vec{q}\,\sigma_A+\omega_{nl}\omega_l\sigma_C)/(\omega_nk_{nl}^2),\nonumber\\
\mathcal{K}_D&=&[\vec{p}\cdot\vec{q}(2\omega_{nl}^2-k_{nl}^2)+2\vec{k}\cdot\vec{p} \,\vec{k}\cdot\vec{q}]\omega_l\sigma_{D}/(\vec{p}^2\omega_n k_{nl}^2),\nonumber\\
\mathcal{K}_{A_5}&=&[(\vec{p}\cdot \vec{q}\, k_{nl}^2+2\vec{k}\cdot \vec{p}\, \vec{k}\cdot\vec{q})\sigma_{A_5}+2\vec{k}\cdot\vec{p}\,\omega_{nl}\omega_l\sigma_{C_5}]/(\vec{p}^2 k_{nl}^2),\nonumber\\
\mathcal{K}_{B_5}&=&-3\sigma_{B_5},\nonumber\\
\mathcal{K}_{C_5}&=&\omega_l\sigma_C/\omega_n+2\omega_{nl}(\vec{k}\cdot\vec{q}\,\sigma_{A_5}+\omega_{nl}\omega_l\sigma_{C_5})/(\omega_nk_{nl}^2),\nonumber\\
\mathcal{K}_{D_5}&=&[\vec{p}\cdot\vec{q}(2\omega_{nl}^2-k_{nl}^2)+2\vec{k}\cdot\vec{p}\, \vec{k}\cdot\vec{q}]\omega_l\sigma_{D_5}/(\vec{p}^2\omega_n k_{nl}^2).
\end{eqnarray}

Then, we can solve Eq.~(\ref{Aetal}) numerically for specific chiral chemical potential $\mu_5$, as well as for the temperature $T$ and normal chemical potential $\mu$. The critical $T_c[\mu_5]$ and $\mu_c[\mu_5]$, which are coordinates of the new ``critical end point'' at a specific $\mu_5$ (CEP[$\mu_5$]), are determined by the scalar susceptibility that is defined in Eq.~(\ref{chim})~\footnote{Here we summarize that, in this work ``CEP'' means the critical end point in the $T-\mu$ plane with $\mu_5=0$, while CEP$_5$ denotes the possible one in the $T-\mu_5$ plane with $\mu=0$~\cite{PhysRevD.84.014011}, and CEP[$\mu_5$] is generally the similar critical end point in the $T-\mu$ plane for a specific $\mu_5$.}.  In this work, we will concentrate on the behavior of the CEP[$\mu_5$], which is expected to be linked to a possible CEP$_5$ in the $T-\mu_5$ plane~\cite{PhysRevD.84.014011}. In Fig.~\ref{tmu5} and Fig.~\ref{mumu5}, we plot the relations between $\mu_5$ and $T_c[\mu_5]$ as well as $\mu_5$ and $\mu_c[\mu_5]$, respectively, which are obtained by seeking the corresponding CEP[$\mu_5$] for different $\mu_5$. Therefore, each point in these two lines means a ``CEP' in the $T-\mu$ plane for the corresponding $\mu_5$.
\begin{figure}
\includegraphics[width=0.45\textwidth]{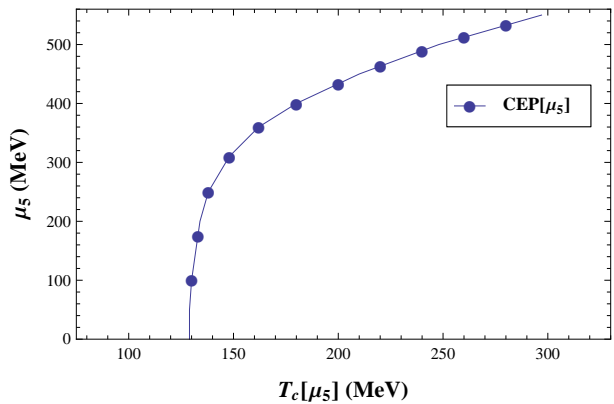}
\caption{The relation between $\mu_5$ and the corresponding $T_c[\mu_5]$ in the $T-\mu$ plane.}\label{tmu5}
\end{figure}
\begin{figure}
\includegraphics[width=0.45\textwidth]{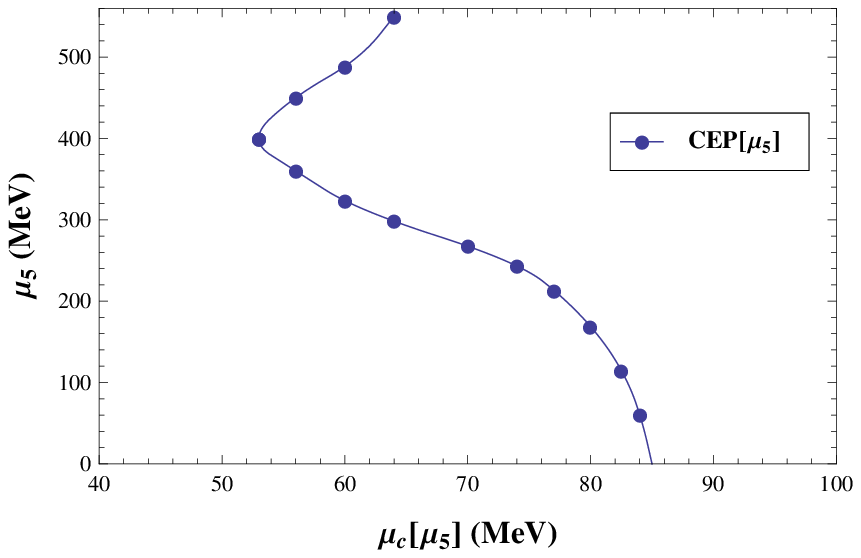}
\caption{The relation between $\mu_5$ and the corresponding $\textcolor{blue}{\mu_c}[\mu_5]$ in the $T-\mu$ plane.}\label{mumu5}
\end{figure}

We can see Fig.~\ref{tmu5} that, $T_c[\mu_5]$ increase slowly when $\mu_5$ is smaller than approximately 300 MeV, and turns to increase quickly for larger $\mu_5$. Nevertheless, the whole increase is smooth and monotonic. The most interesting thing is as shown in Fig.~\ref{mumu5}, that $\mu_c[\mu_5]$ will decrease firstly when $\mu_5$ is smaller than about 400 MeV, and then increase for larger $\mu_5$. The behavior of $\mu_c[\mu_5]$ for different $\mu_5$ is quite different with the previous results from some chiral models, such as Fig.~4 of Ref.~\cite{PhysRevD.84.014011}, that at some critical value of $\mu_5$, $\mu_c[\mu_5]$ will decrease to 0, where a CEP$_5$ is expected to exist. When $\mu_5$ is not very large, the qualitative properties of the results in Fig.~\ref{mumu5} are similar to those from chiral models, but the decrease of $\mu_c[\mu_5]$ is much slower. Our results also indicate there might be no CEP$_5$ in the $T-\mu_5$ plane, and the phase transition in the $T-\mu_5$ plane might be totally crossover when $\mu<50$ MeV,  which then has apparent consistency with the Lattice QCD calculation~\cite{PhysRevLett.107.031601}. Furthermore, our studies may also provide some hints for recent studies related to the chiral chemical potential, such as Ref.~\cite{PhysRevD.90.074009}. Since comparing with the chiral models, DSEs are renormalizable, and include the effects of color confinement as well as DCSB simultaneously, accordingly the DSEs approach is commonly accepted to be a closer theory to QCD itself~\cite{PPNP.33.477,PPNP.45.S1,IJMPE.12.297,PPNP.61.50,PPNP.77.1-69}. In the calculations within the framework of DSEs, people do not need to introduce some annoying parameters, such as the momentum cutoff scale which destroys some basic symmetries of QCD, and then can give more reliable results. In other words, although the chiral models have been found to give reasonable phase diagrams in $T-\mu$ plane, they can not guarantee the robustness when $\mu_5$ acts as an additional scale, and the introducing of $\mu_5$ might make the applicability of the calculations $T+\mu+\mu_5<\Lambda$ ($\Lambda$ is some cutoff scale).

\section{Summary}\label{sum}
Thanks to the complicated non-Abelian feature of Quantum Chromodynamics (QCD) itself, its two fundamental features, namely, dynamical chiral symmetry breaking (DCSB) and quark color confinement, have to be studied phenomenologically through various effective models at present, especially in the most interesting non-perturbative region. In this work, we discuss the chiral phase transition of QCD within the framework of Dyson-Schwinger equations (DSEs), with a chiral chemical potential $\mu_5$ as an additional scale other than the normal temperature $T$ and quark chemical potential $\mu$. We give a basic introduction to the DSEs at finite temperature and nonzero chemical potential as well as an effective model gluon propagator firstly, and then mainly focus on the calculations related to the famous critical end point (CEP) in the $T-\mu$ plane, which is predicted by many model studies, and has caused much interests both in the experimental side (one of the main goals in some high energy physics experiments) and theocratical side. With the help of a scalar susceptibility, that often act as an order parameter of chiral phase transition, we find that there might be no CEP$_5$ in the $T-\mu_5$ plane, which is thought to exist by some chiral model calculations, and the phase transition in the $T-\mu_5$ plane might be totally crossover when $\mu<50$ MeV, which has apparent consistency with the Lattice QCD calculation. DSEs is widely used as well as has been proved to be successful in hadron physics and phase transitions of strongly interacting matters, so that our study may also provide some useful hints to some other studies related to $\mu_5$. Last but not least, we'd like to say that the related issues deserve further studies.

\acknowledgments
This work is supported in part by the National Natural Science Foundation of China (under Grant Nos. 11275097, 11475085, 11265017, and 11247219), the National Basic Research Program of China (under Grant No. 2012CB921504), the Jiangsu Planned Projects for Postdoctoral Research Funds (under Grant No. 1402006C), the National Natural Science Foundation of Jiangsu Province of China (under Grant No. BK20130078), and Guizhou province outstanding youth science and technology talent cultivation object special funds (under Grant No. QKHRZ(2013)28).

\section*{appendix: Structure of the quark propagator}\label{appA}
Using Eq. (\ref{S-1}), after some algebra we find that the quark propagator can be written as (note that here ${\cal F}={\cal F}(\vec{q}^2,\tilde\omega_l^2,\mu_5)$),
\begin{eqnarray}\label{S}
&&S(q_l,\mu_5)=i\not\!\vec{q}\sigma_A+\sigma_B+i\gamma_4\tilde\omega_l\sigma_C+\not\!\vec{q}\gamma_4\tilde\omega_l\sigma_D\nonumber\\
&&+(i\not\!\vec{q}\sigma_{A_5}+\sigma_{B_5}+i\gamma_4\tilde\omega_l\sigma_{C_5}+\not\!\vec{q}\gamma_4\tilde\omega_l\sigma_{D_5})\gamma_5,
\end{eqnarray}
where
\begin{eqnarray}\label{A1etal}
\sigma_A&=&(-At_1+C_5\omega_lt_2)/t_3,\nonumber\\
\sigma_B&=&(Bt_1-D_5\omega_l\vec{q}^2t_2)/t_3,\nonumber\\
\sigma_C&=&(-Ct_1-A_5\vec{q}^2t_2/\omega_l)/t_3,\nonumber\\
\sigma_D&=&(-Dt_1-B_5t_2/\omega_l)/t_3,\nonumber\\
\sigma_{A_5}&=&(A_5t_1-C\omega_lt_2)/t_3,\nonumber\\
\sigma_{B_5}&=&(-B_5t_1+D\omega_l\vec{q}^2t_2)/t_3,\nonumber\\
\sigma_{C_5}&=&(-C_5t_1-A\vec{q}^2t_2/\omega_l)/t_3,\nonumber\\
\sigma_{D_5}&=&(-D_5t_1-Bt_2/\omega_l)/t_3,
\end{eqnarray}
and
\begin{eqnarray}\label{t123}
t_1&=&B^2-B_5^2+(C^2-C_5^2)\omega_l^2\nonumber\\
&&+[A^2-A_5^2+(D^2-D_5^2)\omega_l^2]\vec{q}^2,\nonumber\\
t_2&=&2\omega_l(A_5C+B_5D-AC_5-BD_5),\nonumber\\
t_3&=&t_1^2+\vec{q}^2t_2^2,
\end{eqnarray}

\bibliography{xuss}
\end{document}